\documentclass[12pt]{article}

\usepackage{amssymb}
\usepackage{amsmath}
\usepackage{amscd}
\usepackage{latexsym}
\usepackage{graphicx}

\usepackage{cite}

\newcommand{\be}{\begin{equation}}
\newcommand{\ee}{\end{equation}}

\newcommand{\dlt}{\delta}

\newcommand{\ra}{\rightarrow}

\newcommand{\cA}{{\cal A}}
\newcommand{\cP}{{\cal P}}
\newcommand{\cH}{{\cal H}}

\newcommand{\bt}{\beta}

\newcommand{\al}{\alpha}

\newcommand{\rgl}{\rangle}
\newcommand{\lgl}{\langle}

\begin{document}

\begin{center}

{\Large {\bf Tossing Quantum Coins and Dice} \\ [5mm]

V.I. Yukalov$^{1,2}$}

{\it
$^{1}$Bogolubov Laboratory of Theoretical Physics,
Joint Institute for Nuclear Research, \\
Dubna 141980, Russia \\
E-mail: yukalov@theor.jinr.ru \\ 
and \\
$^{2}$Instituto de Fisica de S\~ao Carlos, Universidade de S\~ao Paulo, \\
CP 369,  S\~ao Carlos 13560-970, S\~ao Paulo, Brazil}

\end{center}

\vskip 3cm

\begin{abstract}
The procedure of tossing quantum coins and dice is described. This case is an
important example of a quantum procedure because it presents a typical framework 
employed in quantum information processing and quantum computing. The emphasis is
on the clarification of the difference between quantum and classical conditional 
probabilities. These probabilities are designed for characterizing different systems, 
either quantum or classical, and they, generally, cannot be reduced to each other. 
Thus the L\"{u}ders probability cannot be treated as a generalization of the 
classical conditional probability. The analogies between quantum theory of 
measurements and quantum decision theory are elucidated.
 
\end{abstract}

\vskip 2cm

{\bf Keywords}: quantum coins; quantum dice; quantum and classical probabilities; 
quantum conditional probability; quantum decision theory

\newpage

\section{Introduction}

Reformulation of classical processes into the quantum language is an interesting and 
important trend in the recent endeavour of applying quantum techniques to the problems 
of quantum information processing, quantum computing, and quantum cognition 
\cite{Williams_1,Nielsen_2,Eisert_3,Vedral_4,Keyl_5,Piotrowski_6,Landsburg_7,Guo_8,
Horodecki_9,Guhne_10,Yukalov_11,Khrennikov_12,Busemeyer_13,Wilde_14,Bagarello_16,
Helland_15,Bagarello_17,Yukalov_18}. Quantum devices provide essentially more operational 
power as compared to their classical counterparts. The widespread point of view is that 
a quantum process is a straightforward generalization of a classical process. However 
the situation is much more complicated. In some cases, it is really a straightforward 
generalization, although in other cases it is just a different process, but not a 
generalization in a strict sense. Quite often, quantum and classical processes are so much 
different that the former cannot be reduced to the latter in the sense that the latter 
is not a particular case of the former. 

The aim of this paper is to illustrate the said above by the example of a quantum
process translating the well known classical tossing of coins and dice to the quantum
language. Since this procedure allows for a clear cut formulation in quantum terms, 
it enjoys three advantages. First, it serves as an explicit example of how a classical 
process can be converted into the corresponding quantum counterpart. Second, it 
demonstrates the difference between quantum and classical procedures. At the same time, 
it explains the principal difference between quantum and classical probabilities, 
especially between conditional probabilities. The last, but not the least, this example 
elucidates the relation between the theory of quantum measurements and quantum decision 
theory. 

The tossing of quantum coins or dice can be realized with different physical objects, for
instance with photons with left and right polarizations or with particle spins having 
several projections. Coin flipping is often treated as a cryptographic problem in which 
a pair of remote distrustful parties must generate a random bit that they agree on 
\cite{Molina_62,Aharon_63}. The present paper considers other issues. One is the problem 
of repeated coin or dice tossing and the relation between classical and quantum conditional 
probabilities. We show that the von Neumann-L\"{u}ders probability cannot be accepted as 
a generalization of the classical conditional probability. The other problem is the 
influence of quantum noise on the results of measurements. Drawing a parallel between 
the mathematical techniques of quantum measurement theory and quantum decision theory 
makes it straightforward to develop a way of taking into account irrational feelings 
imitating quantum random noise of measurements.

\section{Quantum Coins and Dice}

Quantum objects are usually defined in the frame of a Hilbert space $\mathcal{H}$ that 
can be represented as a closed linear envelope over a basis,
\be
\label{1}
 \cH = {\rm span} \{ |\; j\; \rgl \} \;  ,
\ee
where $j$ is a labeling index. Here and in what follows, the Dirac bracket (bra-ket)
notation is used \cite{Dirac_19,Dirac_20}. A quantum system is associated with a 
statistical operator $\hat{\rho}$ which is a semi-positive trace-one operator acting 
on $\cH$. The statistical operator is also called the {\it system quantum state} or simply 
the {\it system state}. The pair $\{\cH,\hat\rho\}$ forms a {\it quantum statistical 
ensemble}. Physical quantities are represented by the algebra of local observables 
$\mathcal{A}$ that is a von Neumann algebra on $\cH$. Von Neumann algebras are 
self-adjoint, strong-operator closed subalgebras (containing the identity operator) 
of the algebra of all bounded operators on a Hilbert space. Observables compose a 
set of averages of the operators from $\mathcal{A}$, termed the {\it normal state} 
that is given by the trace
\be
\label{2}
  \lgl \cA \rgl = {\rm Tr}\;\hat\rho \;\cA \; ,
\ee
where the trace operation is over $\mathcal{H}$. A {\it quantum system} is a quantum 
statistical ensemble and the von Neumann algebra of observables $\{\cH,\hat\rho,\cA\}$. 
An important subclass of $\mathcal{A}$ is the projection ring $\mathcal{P}(\mathcal{A})$ 
of all projection operators belonging to $\mathcal{A}$. 

A subset of projection operators from $\mathcal{P}(\mathcal{A})$, forming an orthomodular 
lattice, composes a projection-valued measure, provided a probability measure is defined 
on this projection lattice. A quantum statistical ensemble and a projection-valued measure 
form a quantum probability space $\{\mathcal{H}, \hat{\rho}, \mathcal{P} \}$. Details 
of the general description of the related constructions can be found in the papers 
\cite{Gudder_21,Cassinelli_22,Goodman_23,Niestegge_24,Bobo_25}. Here we aim at considering 
a concrete case associated with the toss of coins and dice. 

Let the observable of interest be represented by an operator $\hat{A} \in \mathcal{A}$.
The operators of observables are self-adjoint and possess real-valued eigenvalues given 
by the eigenproblem  
\be
\label{3}
 \hat A \; |\;  A_n  \; \rgl  =  A_n \; |\;  A_n  \; \rgl \; .
\ee
The eigenvectors $|A_n\rangle \in \mathcal{H}$ can always be orthonormalized so that
$$
\lgl \; A_m \; |\;  A _n \; \rgl =\dlt_{mn} \;  .
$$
We keep in mind a nondegenerate case because of two reasons. First, this situation 
better suits the studied process of tossing, where each outcome is a simple uniquely 
defined event. Second, the occurrence of degeneracy just slightly complicates the 
consideration without changing conclusions \cite{Yukalov_26}.       
 
The index $n$ enumerates the available alternatives, taking the values $n=1,2,\ldots,N_A$. 
In the application to coins, an event $A_n$, with $n = 1,2$ signifies the occurrence of 
either heads or tails, while in the case of a standard dice, $A_n$, with $n=1,2,3,4,5,6$, 
are the numbers coming up on the sides of a die. There exist also multi-sided dice with $n$ 
varying from $1$ to $120$. The projection-valued measure $\mathcal{P}(\hat{A})$ is 
composed of the projection operators
\be
\label{4}
 \hat P(A_n) = |\;  A_n \; \rgl \lgl \; A_n \; |\;  .
\ee
The operator of an observable can be represented as
\be
\label{5}
 \hat A = \sum_n A_n \hat P(A_n) \;  .
\ee
The probability measure is given by the average
\be
\label{6}
 p(A_n) = \lgl \; \hat P(A_n) \; \rgl = {\rm Tr}\;\hat\rho \;\hat P(A_n) \;
\ee
and enjoys the normalization
\be
\label{7}
 \sum_n \; p(A_n) =  1 \; , \qquad 0\leq p(A_n) \leq 1 \;  .
\ee
With the projection operator (\ref{4}), probability (\ref{6}) takes the form
\be
\label{8}
 p(A_n) = \lgl \; A_n \; | \; \hat\rho\; |\; A_n \; \rgl \;  .
\ee
Note that probability (\ref{6}) is the a priori or predicted probability.

We need to compare result (\ref{8}) with the outcomes of tossing classical coins or 
dice, where we shall denote the classical probability of an event $A_n$ by $f(A_n)$ 
in order to distinguish it from the quantum probability $p(A_n)$. As we know, a fair 
coin or dice give $f(A_n)=1/N_A$. So, generally, the quantum and classical probabilities 
are different. However, it is possible to point out a special case, when the quantum 
probability reduces to the classical one. This is when the system state is pure, so that
$$
\hat\rho = |\; \psi \; \rgl \lgl \;\psi \; | \qquad 
( |\; \psi \; \rgl \in \cH ) \; ,
$$
and has the specially prepared wave function
$$
|\; \psi \; \rgl = \sum_n \; \frac{1}{\sqrt{N_A}} \; |\; A_n \; \rgl \;   .
$$
Under this condition, $p(A_n)=f(A_n)=1/N_A$. 

Since under special explicitly formulated conditions, the quantum probability $p(A_n)$ 
reduces to the classical probability $f(A_n)$, one says that the quantum probability is 
a generalization of the classical probability.

\section{Probability of Repeated Tosses}

The situation is more intricate in the case of repeated tosses of a coin or a die, 
when one tries to define quantum conditional probabilities. The correct consideration 
of repeated measurements requires a description of processes developing in time 
\cite{Yukalov_18}. Then the probability of an event $A_n$ occurring at time $t$ 
writes as
\be
\label{9}
 p(A_n,t) = {\rm Tr}\; \hat\rho(t) \;\hat P(A_n) \;  .
\ee
Dependence of the system state on time is regulated by the law
\be
\label{10}
\hat\rho(t) = \hat U(t,0) \; \hat\rho(0)\; \hat U^+(t,0)  \;  ,
\ee
where $\hat{U}(t,0)$ is the unitary evolution operator describing the evolution of 
the initial state $\hat{\rho}(0)$ at time $t=0$ to a finite time $t>0$. The a priori 
probability of an event $A_n$ in the interval $0 \leq t < t_0$ is given by expression 
(\ref{9}). 

Suppose at the moment of time $t_0$ there appears a new information on the event $A_n$, 
because of which the system starts a new dynamics with the changed initial condition,
\be
\label{11}
\hat\rho(t_0-0) \longmapsto \hat\rho(A_n,t_0) \;   .
\ee
This change is termed {\it state reduction}, although its meaning is just the postulation 
of the new initial condition caused by the received information.
 
The new dynamics for time $t \geq t_0$ is regulated by the evolution law
\be
\label{12}
\hat\rho(A_n,t) = \hat U(t,t_0) \; \hat\rho(A_n,t_0) \; \hat U^+(t,t_0) \;  .
\ee
Therefore the subsequent occurrence of an event $A_m$ at time $t \geq t_0$ happens 
with the probability
\be
\label{13}
p(A_m,t \; | \; A_n,t_0) = {\rm Tr}\; \hat\rho(A_n,t) \;\hat P(A_m) \; .
\ee

The operator $\hat U(t,t_0)$ is the evolution operator corresponding to the considered
quantum system. For instance, a quantum coin can be represented by an electron with the 
spin either up or down. This electron has to be kept inside a quantum system by means of 
trapping potentials and external fields. During the time after a measurement, the state of 
the electron varies depending on the concrete used trapping potentials and applied fields. 
These concrete potentials and fields define the explicit form of the evolution operator.

If the new information on the event $A_n$ consists in the knowledge that at the time 
$t_0$ this event has certainly occurred, then one has the requirement 
\be
\label{14}
 p(A_n,t_0 \; | \; A_n,t_0) = 1 \; ,
\ee
which implies
\be
\label{15}
  {\rm Tr}\; \hat\rho(A_n,t_0) \;\hat P(A_n) =1 \; .
\ee
The solution to the latter equation, suggested by the von Neumann and L\"{u}ders, 
\cite{Neumann_27,Luders_28} is the state
\be
\label{16}
 \hat\rho(A_n,t_0) = 
\frac{\hat P(A_n)\hat\rho(t_0-0)\hat P(A_n)}{ {\rm Tr}\hat\rho(t_0-0)\hat P(A_n)} \; .  
\ee
By the direct substitution, it is straightforward to check that the von Neumann-L\"{u}ders 
state (\ref{16}) is really a solution to equation (\ref{15}). This solution is general in 
the sense of being valid for arbitrary $A_n$ and $\hat{\rho}(t_0 - 0)$, under the initial 
condition change (\ref{11}).  

As an illustration of the state reduction, let us consider the case where before the 
initial condition change (\ref{11}) there was a pure state
$$
\hat\rho(t_0-0) = |\; \psi(t_0)\; \rgl \lgl \; \psi(t_0) \; | \;   .
$$
Then in view of the equalities
$$
 \hat P(A_n)\; \hat\rho(t_0-0)\; \hat P(A_n) = 
|\; \lgl \; A_n \; | \; \psi(t_0) \; \rgl \; |^2 \; \hat P(A_n) \;  , 
\qquad
{\rm Tr} \; \hat\rho(t_0-0)\; \hat P(A_n) = 
|\; \lgl \; A_n \; | \; \psi(t_0) \; \rgl \; |^2 \; ,
$$ 
we get the pure state
$$
 \hat \rho(A_n,t_0) =  |\; A_n \; \rgl \lgl \; A_n \; | \;  .
$$

The probability (\ref{13}) of a subsequent event $A_m$, occurring at a time $t>t_0$,
reads as
$$
p(A_m,t|A_n,t_0) =
\frac{{\rm Tr}\hat U(t,t_0)\hat P(A_n)\hat\rho(t_0-0)\hat P(A_n)\hat U^+(t,t_0)\hat P(A_m)}
{{\rm Tr}\hat\rho(t_0-0)\hat P(A_m)} \;   .
$$
The latter expression depends on the details of the concrete physical system, with 
its trapping potentials and fields, chosen for representing the quantum coin or dice, 
which define the explicit form of the evolution operator. Some simplification of the 
expression can be achieved in the case of nondestructive measurements
\cite{Kampen_36,Shao_37,Braginsky_38,Mozyrsky_39,Yukalov_40,Yukalov_41}, when the 
matrix elements of the evolution operator can be represented as \cite{Yukalov_18}
$$
\lgl \; m \; | \; \hat U(t,t_0)\; | \; n \; \rgl =
\dlt_{mn}\exp\left\{ - i \int_{t_0}^t E_n(t')\; dt' \right\} \;   .
$$    

In quantum theory, one considers successive measurements accomplished immediately 
one after the other, when
$$
\lim_{t\ra t_0+0} \;  \lgl \; m \; | \; \hat U(t,t_0)\; | \; n \; \rgl
= \dlt_{mn} \;  .
$$
It is exactly this case that is usually treated \cite{Neumann_27,Luders_28}, since 
then the consideration does not depend on the details of the evolution operator. 

The probability of observing an event $A_m$ at time $t_0 + 0$, immediately after the 
event $A_n$ has certainly happened at time $t_0$, becomes
\be
\label{17}
p(A_m,t_0+0\; | \; A_n,t_0) = {\rm Tr}\; \hat\rho(A_n,t_0) \;\hat P(A_m) \; ,
\ee
or explicitly
\be
\label{18}
p(A_m,t_0+0\; | \; A_n,t_0) = 
\frac{{\rm Tr}\;\hat\rho(t_0-0)\hat P(A_n)\hat P(A_m)\hat P(A_n)} 
{{\rm Tr}\;\hat\rho(t_0-0)\hat P(A_n)} \;  .
\ee
This expression is termed the L\"{u}ders probability. 

Since $\hat{P}(A_m)$ and $\hat{P}(A_n)$ commute with each other, we have
\be
\label{19}
p(A_m,t_0+0\; | \; A_n,t_0) = 
\frac{{\rm Tr}\;\hat\rho(t_0-0)\hat P(A_m)\hat P(A_n)} 
{{\rm Tr}\;\hat\rho(t_0-0)\hat P(A_n)} \;   .
\ee
This relation reminds us the expression for a classical conditional probability. 
However this impression is misleading, since the left-hand and right-hand sides 
of this relation are defined for different times, but not for the same time as it 
should be for the correct definition of a conditional probability. Really, in the 
left-hand side, we see an a posteriori quantity determined at the time $t_0+0$, 
after the event $A_n$ has happened, while in the right-hand side, we have the a 
priori probabilities ascribed to the time $t_0-0$, before any event has occurred.  

The explicit form of the L\"{u}ders probability can be easily found taking into 
account that
$$
{\rm Tr}\;\hat\rho(t_0-0)\; \hat P(A_n)\; \hat P(A_m)\; \hat P(A_n) =
|\; \lgl \;A_m \; | \; A_n \; \rgl \; |^2 \; 
{\rm Tr}\;\hat\rho(t_0-0)\; \hat P(A_n) \;  .
$$
Therefore the L\"{u}ders probability becomes
\be
\label{20}
p(A_m,t_0+0\; | \; A_n,t_0) = |\; \lgl \;A_m \; | \; A_n \; \rgl \; |^2 \;  ,
\ee
which is a transition probability between the states $|A_m\rangle$ and $|A_n\rangle$. 
Moreover, in view of the property $\langle A_m|A_n \rangle = \delta_{mn}$, we get
\be
\label{21}
 p(A_m,t_0+0\; | \; A_n,t_0) = \dlt_{mn} \; .
\ee
Thus the L\"{u}ders probability for the case of tossing quantum coins or dice is either
one or zero,
$$
p(A_n,t_0+0\; | \; A_n,t_0) = 1 \; ,
$$
\be
\label{22}
p(A_m,t_0+0\; | \; A_n,t_0) = 0 \qquad ( m \neq n) \;  .
\ee

This result is in the drastic difference with the case of tossing classical coins or 
dice, when the conditional probability has the form
\be
\label{23}
 f(A_m\; | \; A_n ) = \frac{f(A_m\cap A_n)}{f(A_n)} \; .
\ee
Keeping in mind that different toss-ups are independent, hence
\be
\label{24}
 f(A_m\cap A_n) = f(A_m) f(A_n) \;  ,
\ee
we come to the value
\be
\label{25}
  f(A_m\; | \; A_n ) = f(A_m) = \frac{1}{N_A} \;  .
\ee

As is evident, the quantum L\"uders probability (\ref{21}) in no way can be reduced 
to the classical conditional probability (\ref{25}). In that sense, the L\"{u}ders 
probability cannot be named a generalization of a classical conditional probability. 
The latter and the L\"{u}ders probability are different quantities describing very 
different systems, a quantum system and a classical system. The L\"{u}ders probability, 
contrary to the classical conditional probability, possesses the following distinct 
properties. 

(i) It is defined for different times, connecting an a posteriori expression in 
the left-hand side with a priori probabilities in the right-hand side of relation 
(\ref{19}).  

(ii) It does not satisfy the statistical independence rule, similar to the classical 
property (\ref{25}). 
   
(iii) It cannot be reduced to the classical expression (\ref{25}). 
    
This result teaches us that, resorting to the L\"{u}ders procedure for the 
interpretation of probabilistic events, one should not forget the true meaning 
behind this procedure. This procedure is designed for describing a two-step 
process of a repeated quantum measurement. The first measurement occurs for an 
arbitrary given initial state of the system and involves no information on the 
future second measurement. While the second measurement is accomplished over 
the quantum state transformed by the first measurement. Any formal operations 
neglecting this true meaning of the L\"{u}ders procedure have no sense.     

The considered simple example of tossing quantum coins or dice demonstrates that 
the classical conditional probabilities and the quantum probabilities defined by 
the L\"uders rule are principally different. Even for commutative operators, when 
the L\"uders probability formally reminds the classical conditional probability, 
their meanings and values remain very distinct. The L\"uders probability is not 
a generalization of the classical conditional probability simply because the former 
cannot be reduced to the latter. And this is not surprising: quantum and classical 
systems have very different properties, so that their characteristics are not obliged 
to identically coincide. The basic difference between quantum and classical cases is 
that a quantum measurement essentially changes the state of the considered quantum 
system, while a measurement in the classical setup has no influence on the system. 

It is important to stress that Eqs. (21) and (22) are in agreement with {\it the 
principle of measurement reproducibility} in quantum theory \cite{Neumann_27} 
according to which: {\it The second measurement, accomplished immediately after 
the first, has to reproduce the result of the latter}. No such principle exists 
for classical systems.

\section{Alternation of Coins and Dice}

It is useful to consider a bit more general case, when the toss of a coin is alternated 
with that of a die, or when two different types of dice are tossed in turn. This means
that, after tossing one object, say a coin, and getting one of the events $A_n$, with 
$n=1,2,\ldots,N_A$, we toss another object, say a die, with possible toss-up events 
$B_k$, where $k=1,2,\ldots,N_B$. 

Similarly to the previous consideration, for the operator of an observable 
$\hat{B}\in\cA$ we have the eigenproblem
\be
\label{26}   
\hat B \; | \; B_k \; \rgl = \hat B_k \; | \; B_k \; \rgl  \; ,
\ee
with the orthonormalized eigenvectors,
$$
 \lgl \; B_k \; | \; B_p \; \rgl = \dlt_{kp} \;  .
$$
The related projection operators are 
\be
\label{27}
\hat P(B_k) = | \; B_k \; \rgl \lgl \; B_k \; | \;   .
\ee

By assumption, in the time interval $0 \leq t < t_0$, the system evolved as in 
Eq. (\ref{10}). At the moment of time $t_0$, the event $A_n$ was observed, so that 
the system started a new dynamics with the initial condition (\ref{16}) for the 
system state. The quantum statistical ensemble, corresponding to the new dynamics, 
is now given by the pair $\{\mathcal{H}, \hat{\rho}(A_n,t)\}$. The projection-valued 
measure $\mathcal{P}(\hat{B})$ consists of the projection operators (\ref{27}). The 
quantum probability space, characterizing measurements with respect to the observable 
$\hat{B}$, is formed by the new statistical ensemble and the projection-valued measure
$\mathcal{P}(\hat{B})$, that is, this probability space is
$\{\cH,\hat\rho(A_n,t),\mathcal{P}(\hat{B})\}$. In that way, the probability
of observing an event $B_k$, after the event $A_n$ has been certainly observed, is
\be
\label{28}
  p(B_k,t \; | \; A_n,t_0) = {\rm Tr}\; \hat\rho(A_n,t) \; \hat P(B_k) \; .
\ee
Accomplishing the observation for $\hat{B}$, immediately after $A_n$ was observed at 
the moment of time $t_0$, defines the probability
\be
\label{29}
  p(B_k,t_0+0 \; | \; A_n,t_0) = 
{\rm Tr}\; \hat\rho(A_n,t_0) \; \hat P(B_k) \; ,
\ee
which is the L\"{u}ders probability
\be
\label{30}
  p(B_k,t_0+0 \; | \; A_n,t_0) = 
\frac{{\rm Tr}\hat\rho(t_0-0)\hat P(A_n)\hat P(B_k)\hat P(A_n)}
{{\rm Tr}\hat\rho(t_0-0)\hat P(A_n)} \; .
\ee
Substituting here the related projectors yields
\be
\label{31}
 p(B_k,t_0+0 \; | \; A_n,t_0) = |\; \lgl\; B_k\; | \; A_n \; \rgl\; |^2 \; .
\ee

The obtained L\"{u}ders probability has the meaning of the transition probability 
from a quantum state $|A_n\rangle$ to the state $|B_k\rangle$. The operators $\hat{A}$ 
and $\hat{B}$ do not necessarily commute with each other. Their associated eigenvectors, 
in general, are not orthogonal. Hence, depending on the final and initial quantum states, 
the L\"uders probability (\ref{31}) can be any number between $0$ and $1$. The transition 
probability is symmetric with respect to the interchange of the initial and final states, 
such that
\be
\label{32}
 p(B_k,t_0+0 \; | \; A_n,t_0) = p(A_n,t_0+0 \; | \; B_k,t_0) \; .
\ee
Commuting operators enjoy the same eigenvectors, hence for commuting operators the 
probability (\ref{32}) reduces to $\delta_{nk}$. 
 
The corresponding classical conditional probability
\be
\label{33}
 f(B_k \; | \; A_n) =   f(B_k) = \frac{1}{N_B} 
\ee
does not depend on the initial quantum state. Also, the classical conditional probability 
is not symmetric, since
\be
\label{34}
f(A_n \; | \; B_k) =   f(A_n) = \frac{1}{N_A} \;  ,   
\ee
where $N_A$ does not equal $N_B$. 
 
Thus again we come to the conclusion that the L\"{u}ders probability cannot be reduced 
to the classical conditional probability, hence it cannot be named a generalization 
for the latter, in agreement with the earlier results \cite{Yukalov_26,Yukalov_29}. 
These probabilities are both correct in the regions of their applicability. But they 
are principally different, being designed for very different systems, either quantum or 
classical. A single quantum measurement can be reduced to a single classical measurement, 
but several consecutive quantum measurements, in the frame of the standard quantum-mechanical 
approach, are not reducible to consecutive classical measurements. The root of this basic 
difference lies in the fact that a quantum measurement essentially changes the system state, 
producing the so-called state reduction, while a classical measurement has no influence 
on the studied system.

\section{Degenerate Coins or Dice}

As is shown in the previous section, the L\"{u}ders probability is symmetric with 
respect to the interchange of events, provided the quantum coins and dice are 
described by non-degenerate quantum models. Such a reciprocal symmetry, generally, 
does not happen for the classical conditional probability, as well as contradicts 
experimental observations \cite{Boyer_30,Boyer_31}. The reciprocal symmetry can be 
broken when the considered objects are characterized by degenerate models. 

The quantum degeneracy implies that some eigenvalues of the studied operator are 
associated with several eigenvectors. Let us illustrate the degenerate case for an 
operator $\hat{B}$ for which the eigenproblem is degenerate, when
\be
\label{35}
 \hat B\; |\; B_{k\al} \; \rgl = \hat B_k \; |\; B_{k\al} \; \rgl \qquad
(\al = 1,2,\ldots ) \; .   
\ee
The eigenvectors corresponding to the same eigenvalue can be orthonormalized by means 
of the Gram-Schmidt orthogonalization, so that
$$
 \lgl \; \hat B_{k\al} \; |\; B_{k\bt}\; \rgl =\dlt_{\al\bt} \;  .
$$
The degenerate projection operator takes the form
\be
\label{36}
 \hat P(B_k) = \sum_{\al} \hat P(B_{k\al} ) \;  ,
\ee
where
$$
\hat P(B_{k\al}) =  |\; B_{k\al} \; \rgl \lgl \; \hat B_{k\al} \; | \; . 
$$

The L\"uders probability of a degenerate event $B_k$, occurring after a non-degenerate
event $A_n$ has happened, is
\be
\label{37}
 p(B_k,t_0+0 \; | \; A_n,t_0) = 
\sum_\al |\; \lgl \; B_{k\al} \; | \; A_n \; \rgl \; |^2 \;  .
\ee
Reversing the order of the events yields
\be
\label{38}
p(A_n,t_0+0 \; | \; B_k,t_0) = 
\frac{\sum_{\al\bt} \lgl\; B_{k\al}\;|\;\hat\rho(t_0-0)\;|\;B_{k\bt}\;\rgl 
\lgl\; B_{k\bt}\;|\; A_n\; \rgl \lgl\; A_n\;|\;B_{k\al}\;\rgl}
{\sum_\al \lgl\; B_{k\al}\;|\;\hat\rho(t_0-0)\;|\; B_{k\al}\;\rgl } \; .
\ee
As is clear, generally, there is no reciprocal symmetry for these probabilities.

However, breaking the reciprocal symmetry in no way makes it clear how the L\"{u}ders 
probabilities could be reduced to the classical conditional probabilities. The use of
degenerate models brings the following problems.

First, the meaning of degeneracy for an event not always can be defined. For instance, 
how a coin could be degenerate? There is too much of ambiguity in possible 
interpretations of the degeneracy and in its quantification.

Second, probabilities (\ref{33}) and (\ref{34}) are defined solely by the properties 
of the last events, while probabilities (\ref{37}) and (\ref{38}) depend on both 
conditional and conditioned events. 

Third, trying to reduce a quantum object to a classical one by introducing even more 
quantum features looks absurd. 

Finally, if there exists at least one example where the quantum L\"{u}ders probability 
cannot in principle be reduced to a classical conditional probability, then the former 
cannot be treated as the generalization of the latter. 

It is necessary to accept that the L\"{u}ders probability and the classical conditional 
probability are basically different notions. No one of them is a generalization of the 
other. Each of them allows for a correct mathematical definition and is valid for 
different classes of problems, one for quantum, the other for classical systems. The 
origin of the principal difference is the state reduction in the case of quantum theory, 
but its absence in classical theory. The quantum state reduction does not exist for 
classical systems. Moreover, the principally quantum notion of state reduction cannot be 
translated into the classical language.

\section{State Reduction in Decision Theory}

Theory of quantum measurements can be straightforwardly reinterpreted as quantum decision
theory \cite{Yukalov_26,Yukalov_29,Yukalov_64}. However, it seems that the above analysis 
puts doubt on the possibility of using quantum techniques in decision theory because it 
looks quite unreasonable to imagine that the state of a decision maker, after a decision 
has been made, completely collapses to the state describing just that particular decision. 
Then, after each taken decision, a human subject would be drastically changing, reminding 
not a self-consistent personality but a kind of a weather vane spontaneously turning after 
each decision. 

The problem lies in the fact that in the standard picture, considered above, only the 
space of alternatives has been taken into account. However any observation consists
not merely of a state of alternatives, but involves as well measuring instruments, 
including observers \cite{Yukalov_18}. In cognitive sciences, the role of measuring 
instruments plus an observer is played by a decision maker, with the related 
{\it subject space of mind} that can be represented by a Hilbert space
\be
\label{39}
 \cH_S ={\rm span}\{\; | \; \al\; \rgl \; \} \;  .
\ee
This space characterizes the invariant features of a subject, actually making him/her 
an individual with more or less invariant properties. When considering the process of 
decision making with respect to the set of alternatives $\{A_n\}$, the subject space 
of mind has to be extended by including the {\it state of alternatives}
\be
\label{40}
 \cH_A ={\rm span}\{\; | \; A_n \; \rgl \; \} \;   .
\ee
Thus the total system {\it decision space} becomes
\be
\label{41}
\cH = \cH_S \bigotimes \cH_A \;  .
\ee
The system state is given by a statistical operator $\hat{\rho}(t)$ acting on the 
decision space (\ref{41}).

To explain better the difference of the subject space of mind $\cH_S$ and the space of 
alternatives $\cH_A$, let us take the situation typical of tossing coins or dice. The 
space of alternatives describes the possible outcomes of tossing. An event $A_n$ of 
getting one of the outcomes is represented by the vector $|A_n\rangle$ in the space of 
alternatives. However, the space of outcomes of tossing does not constitute the whole
player who possesses a variety of other interests, feelings, and planned actions that
characterize the player as a human individual. All these additional attributes describing
the player as an individual can be represented by vectors in the subject space of mind. 

When an event $A_n$, corresponding to the observation of the related outcome of tossing, 
occurs, the state in the space of alternatives reduces. For instance, as is explained in 
Sec. 3, if the state before the observation were pure, say 
$|\psi(t_0)\rangle \langle \psi(t_0)|$, after the observation it reduces to the state
$|A_n\rangle \langle A_n|$. However this does not mean that the whole player as a separate
personality has suffered from the reduction to another personality. This only implies
the reduction in the space of the considered alternatives $\cH_A$, but not in the subject
space of mind $\cH_S$ that can stay intact. The player remains the same person almost 
independently of the tossing outcome, although in some life situations the result of 
tossing can influence the state of the player, for instance changing his/her mood or
influencing the planned actions. Vice versa, the subjective feelings and emotions of a 
player can influence the results of tossing. The interrelation between the subject space 
of mind and the space of alternatives will be more in detail explained in Section 10.

The process of tossing coins or dice is characterized by the probability $p(A_n)$ of 
observing an outcome $A_n$. In the frequentist picture, this is equivalent to saying that 
the fraction of people who get the outcome $A_n$ is $p(A_n)$. This probability is
\be
\label{42}
 p(A_n,t) = {\rm Tr}\; \hat\rho(t) \; \hat P(A_n) = 
{\rm Tr}_A\; \hat\rho_A(t) \; \hat P(A_n) \; ,
\ee
with the state in the space of alternatives 
\be
\label{43}
\hat\rho_A(t) = 
{\rm Tr}_S\; \hat\rho(t) \equiv {\rm Tr}_{\cH_S}\; \hat\rho(t) \; .
\ee
The subject state of mind is
\be
\label{44}
 \hat\rho_S(t) = 
{\rm Tr}_A\; \hat\rho(t) \equiv {\rm Tr}_{\cH_A}\; \hat\rho(t)\; .
\ee
         
The certain occurrence of an event $A_n$ at time $t_0$ implies the condition
\be
\label{45}
  p(A_n,t_0+0 \; | \; A_n,t_0) =  
{\rm Tr}_A \; \hat\rho_A(A_n,t_0) \; \hat P(A_n) =  1
\ee
and the reduction 
\be
\label{46}
\hat\rho_A(t_0-0)   \longmapsto \hat\rho_A(A_n,t_0)
\ee
of the state of alternatives to the state
\be
\label{47}
 \hat\rho_A(A_n,t_0) = \frac{\hat P(A_n)\;\hat\rho_A(t_0-0)\;\hat P(A_n)}
{ {\rm Tr}_A\;\hat\rho_A(t_0-0)\;\hat P(A_n) } \; .
\ee

In this way, the state reduction happens only in the space of alternatives (\ref{40}),
while there is no reduction in the subject space (\ref{39}). Therefore the individual 
invariance of the decision maker is left intact. To better illustrate this, let us 
consider the system state in the form
\be
\label{48}
\hat\rho(t) = \hat\rho_S(t) \bigotimes \hat\rho_A(t) \;  ,
\ee
which implies the absence of correlations between the space of alternatives and the 
subject space. Then the reduction of the state of alternatives (\ref{46}) results in 
the reduction of the total system state
\be
\label{49}
\hat\rho(t_0-0)   \longmapsto \rho_S(t_0) \bigotimes \hat\rho_A(A_n,t_0) \; .
\ee
As we see, the state reduction touches only the state of alternatives, while the 
subject state is left invariant. 

When the state reduction occurs only in the space of alternatives, but the decision 
maker, as an individual, remains invariant, this means that the total system state 
changes not much, thus preserving a decision maker as a personality. Hence this 
removes contradictions in applying quantum techniques to cognitive sciences.         

If a subject decides with respect to the same set of alternatives twice, so that the 
second decision follows immediately after the first one, then it is not surprising if 
the decision is the same. Really, we have
\be
\label{50}
  p(A_m,t_0+0 \; | \; A_n,t_0) =  
{\rm Tr}_A\; \hat\rho_A(A_n,t_0)\hat P(A_m) = \dlt_{mn} \; .
\ee
The quantum state reduction, leading to this result, does not contradict realistic 
cognitive processes. This is because it is not the subject who collapses but the 
reduction happens only in the space of alternatives, while the subject, as an 
individual, stays invariant.

\section{Correlated and Separated Measurements}

Although expression (\ref{50}) finds a reasonable interpretation in the frame of 
decision making, nevertheless it is not clear whether there exist such conditions 
when formula (\ref{50}) could be turned into (\ref{25}), or more generally, when a 
quantum conditional probability of two events could be reduced to the classical form 
(\ref{33}) and (\ref{34}). At this point, we again have to recollect that measurements 
or decision processes develop in time \cite{Yukalov_18,Yukalov_26,Yukalov_29}. Let 
one measurement (decision) be accomplished at time $t_0$. Before another measurement 
(decision) could be accomplished, some time necessarily elapses, when the first 
measurement is being fixed, the system evolves, suffers from the disturbance of 
measuring devices, from the influence of surrounding, and from the actions of 
observers preparing another measurement. Let us denote this total time of evolution 
and preparation as $\tau$. So, in reality, two measurements (decisions) are always 
separated from each other by a finite time $\tau$ that we can name the {\it separation 
time}. When one talks about a second measurement (decision) occurring immediately after 
the first measurement, this just implies that the separation time $\tau$ is in some 
sense small. 

To give a meaning to the characteristic ``small", we need to compare $\tau$ with 
a typical time of system variation. This typical time is the {\it relaxation time} 
$t_{rel}$ during which the system, being perturbed, returns to its original state. 
The role of perturbation is played by the process of taking a decision or 
accomplishing a measurement. 

There can happen two qualitatively different situations, when the separation time 
is much smaller than the relaxation time $(\tau \ll t_{rel})$ and the opposite case, 
when $\tau \gg t_{rel}$. If the separation time is much smaller than the relaxation 
time, then the process of the first measurement (decision) cannot be separated from 
the second one. In that case both measurements (decisions) have to be treated, 
actually, as one double strongly correlated measurement consisting of two parts. 
Exactly this situation has been assumed in the previous sections when considering 
the occurrence of a second event immediately after the first one. The opposite case, 
when the separation time is much longer than the relaxation time, can be treated 
as two separate measurements.   

The case, where $\tau \gg t_{rel}$, can be treated in two ways. One way is by 
invoking the exact evolution \cite{Yukalov_18} of the state $\hat{\rho}(t)$, with 
taking into account that for a large separation time, the perturbation, imposed on 
the subject by the first measurement (decision), fades out. Then the probability of 
observing an event $B_k$ at the time $t_0 + \tau$ is not much influenced by the 
measurement occurred at $t_0$. In that case,
\be
\label{51}
 p(B_k,t_0+\tau \; | \; A_n,t_0) \cong p(B_k,t_0+\tau) \;  .
\ee     

The other equivalent way is based on the channel-state duality 
\cite{Yukalov_26,Choi_32,Jamiolkowski_33} and can be realized as follows. Suppose 
at the moment of time $t_0$, one measures (takes a decision) with respect to an 
observable $\hat{A}$ and, after a separation time $\tau$, one attempts performing 
a second measurement with respect to $\hat{B}$. The separation time is assumed to 
be larger than the typical relaxation time, $\tau \gg t_{rel}$. In this situation, 
we have two separate measurements. Then the decision space 
\be
\label{52}
 \cH = \cH_S \bigotimes \cH_A \bigotimes \cH_B 
\ee
can be represented as composed of the subject space (\ref{39}), the space (\ref{40}) 
of alternatives $A_n$, and the space of alternatives $B_k$,
\be
\label{53}
 \cH_B  = {\rm span} \; \{ \; |\; B_k \; \rgl\; \} \; .
\ee

For $t<t_0$, the probability of $A_n$ is given by expression (\ref{42}) acting on 
$\cH_A$, where  
\be
\label{54}
 \hat\rho_A(t) = {\rm Tr}_{SB} \; \hat\rho(t) \; ,
\ee
and the trace operation is over $\cH_S\bigotimes\cH_B$. 

Similarly, we can define the state of alternatives $B_k$, acting on $\cH_B$,
\be
\label{55}
  \hat\rho_B(t) = {\rm Tr}_{SA} \; \hat\rho(t) \; ,
\ee
with the trace over $\cH_S\bigotimes\cH_A$, and the subject state acting on $\cH_S$,
\be
\label{56}
 \hat\rho_S(t) = {\rm Tr}_{AB} \; \hat\rho(t) \; ,
\ee
with the trace over $\mathcal{H_A} \bigotimes \mathcal{H_B}$. 

If at the moment of time $t_0$ an event $A_n$ has certainly happened, then the state 
in the space of alternatives $\mathcal{H_A}$ is reduced according to the rule
\be
\label{57} 
\hat\rho_A(t_0-0) =  {\rm Tr}_{SB} \; \hat\rho(t_0-0) \longmapsto
 {\rm Tr}_{SB} \; \hat\rho(A_n,t_0) = \hat\rho_A(A_n,t_0) \; .
\ee
The conditional probability of observing an event $B_k$, after $A_n$ has been observed,
becomes
\be
\label{58}
 p(B_k,t_0+\tau \; | \; A_n,t_0) =  
{\rm Tr} \; \hat\rho(A_n,t_0+\tau)\; \hat P(B_k) =
{\rm Tr}_B \; \hat\rho_B(t_0+\tau)\; \hat P(B_k) \; .
\ee
If the order of events is reversed, then we have the conditional probability
\be
\label{59}
 p(A_n,t_0+\tau \; | \; B_k,t_0) =  
{\rm Tr} \; \hat\rho(B_k,t_0+\tau)\; \hat P(A_n) =
{\rm Tr}_A \; \hat\rho_A(t_0+\tau)\; \hat P(A_n) \;   .
\ee

Equations (\ref{58}) and (\ref{59}) can be represented as
\be
\label{60}
p(B_k,t_0+\tau \; | \; A_n,t_0) = p(B_k,t_0+\tau) \; , 
\qquad
p(A_n,t_0+\tau \; | \; B_k,t_0) = p(A_n,t_0+\tau) \;  ,
\ee
which have the form of the classical conditional probabilities (\ref{33}) and 
(\ref{34}) describing the probability of tossing coins or dice. In this way, 
the reduction of the quantum conditional probabilities to the related classical 
probabilities of tossing is possible, provided the quantum conditional probabilities 
describe weakly correlated events. This becomes possible due to the consideration of 
the process in the composite space (\ref{52}) characterizing the quasi-independence 
of separate events. Strictly speaking, the interdependence of the events can enter 
through the common state $\hat{\rho}(t)$. The events turn into completely independent 
when
$$
 \hat\rho(t) = 
\hat\rho_S(t) \bigotimes \hat\rho_A(t) \bigotimes \hat\rho_B(t) \; .
$$
The event independence is caused by a long separation time $\tau$, during which the 
correlation between different measurements disappears.   

Thus, for a very short separation time $\tau \ll t_{rel}$, two measurements are strongly 
correlated and have to be treated as a single twin measurement realized in the space 
(\ref{41}), while for a large separation time $\tau \gg t_{rel}$, the measurements are
weakly correlated becoming quasi-independent and characterized by the space (\ref{52}).
In the intermediate case, it is necessary to study the overall evolution of the system
state in time \cite{Yukalov_18}.    

Now we are in a position of summarizing why the L\"{u}ders probabilities (\ref{21})
or (\ref{22}) drastically differ from the classical conditional probabilities (\ref{25})
and why the L\"{u}ders probability (\ref{31}) contradicts the classical conditional 
probabilities (\ref{33}) and (\ref{34}). This happens when one forgets that the correct
description of either measurements or of decision processes requires the consideration 
of temporal evolution. The evolution can be qualitatively different depending on the 
interrelation between the time $\tau$, separating two different measurements (decisions),
and the relaxation time $t_{rel}$, during which the perturbations, caused by the 
measurement (decision) process, attenuate.      

The L\"{u}ders probability is applicable to the case of two consecutive measurements 
(decisions) separated in time by an interval $\tau \ll t_{rel}$, while in the formulation 
of classical conditional probabilities one usually assumes that $\tau$ is either greater
or comparable with $t_{rel}$. For instance, the implementation of two coin tosses
requires quite sufficient time between them, because of which the classical tosses become 
independent of each other. Contrary to this, imaginary quantum coins or dice can be
tossed one after another, assumed to be separated by an arbitrarily short time. Actually,
quantum coins and dice are nothing but quantum bits of information that can be realized
with different quantum objects, such as spins, photons, and atoms \cite{Yukalov_65}.

\section{General Dynamics of Measurements}

In the previous section, it is explained that the properties of quantum probability
are different, depending on the relation between the separation and relaxation times. 
Now we summarize the general description of the measurement dynamics for an arbitrary
interrelation between these typical times. 

As has been mentioned above, the mathematics of realizing measurement procedures, event 
observation, and decision making are practically the same \cite{Yukalov_34,Yukalov_35}, 
because of which it is admissible to use interchangeably either the terminology from 
quantum measurement theory or from quantum decision theory. In the interpretation of 
occurring events, the chance of getting an alternative $A_n$ in a measurement, of course, 
has a different meaning as compared to the procedure of deciding to choose an alternative 
$A_n$. However in both the cases, the mathematics of describing the process are similar.
Thus in measurements, we get an alternative $A_n$ with a probability $p(A_n)$, and in 
decision making, an alternative $A_n$ is chosen with a probability $p(A_n)$. In other 
words, one can say that in measurements, an alternative $A_n$ is found in the fraction
$p(A_n)$ of the realized measurements. While in decision theory, $p(A_n)$ is the fraction
of subjects choosing the alternative $A_n$. In that way, the procedure of describing 
quantum measurements is mathematically similar to the process of decision making 
\cite{Yukalov_18,Yukalov_26,Yukalov_29,Yukalov_34,Yukalov_36}. 

A correct definition of quantum probabilities requires, first of all, to introduce a  
quantum probability space. Generally, this space can be different for the varied stages 
of dynamics, if the latter are interrupted by measurements. Suppose, at the first stage,
lasting from $t=0$ till the time $t_0$, when an event is observed, we are interested in 
studying the probabilities related to the operational variable $\hat{A}$. The decision 
space 
\be
\label{61}
\cH = \cH_S \bigotimes  \cH_A 
\ee
is composed of the subject space $\mathcal{H}_S$ and the space $\mathcal{H}_A$ of 
alternatives $\{A_n\}$. The system state (statistical operator) evolves according to 
the rule
\be
\label{62}
 \hat\rho(t) = \hat U(t,0) \; \hat\rho(0) \; \hat U^+(t,0) \; .
\ee
The operator measure $\mathcal{P}(\hat{A})$ is formed of the projection operators 
$\hat{P}(A_n)$. The quantum probability space is
\be
\label{63}
\{ \cH, \; \hat\rho(t),\; \cP(\hat A)\} \qquad ( 0 \leq t \leq t_0 ) \; .
\ee
The probability measure is defined through the probabilities
\be
\label{64}
 p(A_n,t) = {\rm Tr}\; \hat\rho(t) \; \hat P(A_n) =   
{\rm Tr}_A\; \hat\rho_A(t) \; \hat P(A_n) \; .
\ee
  
At the moment of time $t_0$, the dynamics is interrupted by a measurement revealing 
the occurrence of an event $A_n$. This implies the state reduction
\be
\label{65}
  \hat\rho(t_0-0) \longmapsto \hat\rho(A_n,t_0) \; ,
\ee
such that the projection of the total decision state to the space of alternatives
\be
\label{66}
\hat\rho_A(t_0-0) =  {\rm Tr}_S\; \hat\rho(t_0-0) \longmapsto \hat\rho_A(A_n,t_0)
\ee
acquires the L\"uders form (\ref{47}). The occurrence of the event $A_n$ at $t_0$
implies that 
\be
\label{67}
 {\rm Tr}\; \hat\rho(A_n,t_0) \; \hat P(A_n) =  
 {\rm Tr}_A\; \hat\rho_A(A_n,t_0) \; \hat P(A_n) = 1 \; .
\ee
This equality defines the state $\hat{\rho}_A(A_n,t_0)$ as the L\"{u}ders state. 
The state $\hat{\rho}(A_n,t_0)$ serves as an initial state for the next stage of 
evolution.

The next stage lasts after the time of the measurement $t>t_0$. Now the evolution of 
the system state starts from the new initial condition $\hat{\rho}(A_n,t_0)$, so that
\be
\label{68}
\hat\rho(A_n,t) = 
\hat U(t,t_0) \; \hat\rho(A_n,t_0) \; \hat U^+(t,t_0) \;  .
\ee
At this stage, we are interested in the a priori probabilities of the events $B_k$. 
For the case of tossing, an event $B_k$ can be the result of tossing either the same coin 
or a die or the result of tossing another die. 

Then the operator measure $\mathcal{P}(\hat{B})$ is composed of the projection 
operators $\hat{P}(B_k)$. The quantum probability space becomes
\be
\label{69}
\{ \cH, \; \hat\rho(A_n,t),\; \cP(\hat B) \} \qquad ( t > t_0 ) \;  .
\ee
For this probability space, the a priori probability of observing an event $B_k$ 
at an arbitrary time $t$, 
\be
\label{70}
 p(B_k,t \; | \; A_n,t_0) =  
{\rm Tr} \; \hat\rho(A_n,t)\; \hat P(B_k) \; ,
\ee
plays the role of the quantum conditional probability of observing $B_k$ after 
the event $A_n$ has certainly been observed at time $t_0$.     

As is explained in the previous section, if the dependence of the state 
$\hat{\rho}(A_n,t)$ on the measuring perturbation at the time $t_0$ fades out 
with increasing $t$, so that $t-t_0=\tau\gg t_{rel}$, and this state tends to 
$\hat{\rho}(t)$, then the events $A_n$ and $B_k$ become weakly correlated and we 
verge to the case of independent events (\ref{60}).   

This consideration shows that, generally, the quantum conditional probability 
cannot be represented as in expression (\ref{23}) for the classical conditional 
probability. This can happen only in the limiting case of independent events, when 
the time separation between the events is much larger than the relaxation time.

\section{Synchronous Measurements and Events}

Till now, we have studied conditional probabilities for events occurring at different 
times. A natural question arises: Is it possible to accomplish several measurements 
or observe several events synchronously at the same time? Here we keep in mind 
nonrelativistic events, when the notion of synchronously occurring events or measurements
has sense. In the relativistic case we should use the notion of spacelike separated 
measurements or events. If one considers a realistic system and a realistic 
measurement procedure, then one has to accept that the simultaneous measurement of 
different observables can be realized only by different measuring devices and,
generally, in different parts of the system. For instance, keeping in mind coin and dice 
tossing, we have to agree that a simultaneous toss of either two coins, or a coin and a 
die, or two different dice is feasible only in separate spatial locations. Accepting that 
two measurements are done in two different parts of a system, we shall denote these parts 
as the location $A$ and location $B$.   

In this section, we omit, for the simplicity of notations, the subject space. If 
necessary, it can be easily included back, which does not influence the results. In 
that case, if we measure two observables represented by the operators $\hat{A}$ and 
$\hat{B}$, the process should be described in the composite space of alternatives
\be
\label{71}
 \cH_{AB} = \cH_A \bigotimes \cH_B \; .
\ee
The operators $\hat{A}$ and $\hat{B}$ can be either commuting or not commuting. The 
system state $\rho_{AB}(t)$, acting on this space, can be considered as the state with 
the traced out subject-space degrees of freedom. 

The quantum probability space is
\be
\label{72}
\{ \cH_{AB} , \; \hat\rho_{AB}(t) , \; \cP(\hat A) \bigotimes \cP(\hat B) \} \; .
\ee
Then it is easy to define the probabilities of separate events occurring in the 
corresponding locations
$$
p(A_n,t) = {\rm Tr}_{AB}\; \hat\rho_{AB}(t) \; \hat P(A_n) =
{\rm Tr}_A \; \hat\rho_A(t) \; \hat P(A_n) \;  , 
$$
\be
\label{73}
p(B_k,t) = {\rm Tr}_{AB}\; \hat\rho_{AB}(t) \; \hat P(B_k) =
{\rm Tr}_B \; \hat\rho_B(t) \; \hat P(B_k) \;  ,
\ee
where
$$
 \hat\rho_A(t) = {\rm Tr}_B \; \hat\rho_{AB}(t) \; , \qquad
 \hat\rho_B(t) = {\rm Tr}_A \; \hat\rho_{AB}(t) \;   .
$$

It is also straightforward to introduce the joint probability of two events 
\cite{Yukalov_26,Yukalov_29} occurring at a time $t$,
\be
\label{74}
p(A_nB_k,t) = 
{\rm Tr}_{AB} \; \hat\rho_{AB}(t) \; \hat P(A_n) \bigotimes \hat P(B_k) \;  .
\ee
Notice that the joint probability is symmetric with respect to the event interchange,
\be
\label{75}
 p(A_nB_k,t) = p(B_kA_n,t) \;  .
\ee
The joint probability satisfies the standard properties
\be
\label{76}
 p(A_n,t) = \sum_k p(A_nB_k,t) \; , \qquad p(B_k,t) = \sum_n p(A_nB_k,t) \; .
\ee

The conditional probability of two events, occurring at different locations, can be 
introduced by the usual formula
\be
\label{77}
 p(A_n|B_k,t) \equiv \frac{p(A_nB_k,t)}{p(B_k,t)} \; .
\ee
This conditional probability enjoys the same properties as the classical conditional
probability,
\be
\label{78}
 \sum_n p(A_n|B_k,t) = \sum_k p(B_k|A_n,t) = 1 \; .
\ee
Probability (\ref{77}) is not symmetric with respect to the event interchange, since
\be
\label{79}
 p(B_k|A_n,t) = \frac{p(B_kA_n,t)}{p(A_n,t)} \; .
\ee

The locations $A$ and $B$, in general, are correlated, because of which the state 
$\hat{\rho}_{AB}(t)$ does not have the form of a tensor product. In the particular case, 
when the locations are not correlated, and the events are independent, so that the 
state factorizes,
\be
\label{80}
\hat\rho_{AB}(t) = \hat\rho_A(t) \bigotimes \hat\rho_B(t)  \;  ,
\ee
then 
$$
p(A_nB_k,t) = p(A_n,t) p(B_k,t) \;  ,
$$
and we again get the standard property of conditional probabilities
\be
\label{81}
p(A_n|B_k,t) = p(A_n,t) \; , \qquad  p(B_k|A_n,t) = p(B_k,t) \; .
\ee

In this way, the quantum conditional probability (\ref{77}) can be accepted as a 
generalization of the classical conditional probability.

\section{Quantum Decision Theory} 

The main aim of the present paper has been the detailed analysis of the differences 
between quantum and classical probabilities. To be more illustrative, the case of
tossing coins and dice in both variants, quantum and classical, has been treated. 
The thorough comparison has shown that, strictly speaking, quantum and classical
probabilities are different entities. The quantum probability, such as the L\"{u}ders
probability, cannot be considered as a generalization of the classical conditional 
probability, since the former cannot be reduced to the latter. These are just different 
entities designed for quite differing systems, either quantum or classical. So to speak:
to each its own. 

However for a curious reader, it is difficult to avoid some questions as follows. It 
is clear and well known that quantum probabilities are successfully applied to quantum 
systems. But could they be applied to real-world problems of cognitive science? For 
instance, could quantum probabilities be used in decision making that is so widely 
employed in and so important for many fields of human activity, such as economics, 
finances, managing, and so on? When are quantum probabilities especially useful, or 
needed at all? Which are the types of problems where they could be applied? The answer 
to these questions is given in the present section. 

In addition, explaining when and where quantum probabilities could be involved in 
cognitive sciences, it will become more clear the importance of considering the 
subject space of mind introduced in Sec. 6. There are, of course, quite simple cases, 
where there is no any need of invoking quantum techniques for real-life classical 
problems. Thus, describing the toss of a classical coin does not require the use of 
quantum terminology. Vice versa, dealing with a quantum analog of coin tossing, e.g. 
when observing a spin of a particle, the standard quantum techniques are sufficient. 
However, the use of quantum probabilities in cognitive sciences can be useful in 
analyzing rather complicated situations having to do with some kind of dual influence 
on the decision processes. For this purpose, it is necessary to deviate from the 
trivial case of coin tossing and to turn to essentially more complicated and delicate 
issues of decision making. 

To grasp the feeling why quantum techniques could be useful in decision making, one 
has to remember one of the most important points of quantum theory incorporating the 
so-called particle-wave duality. Then, intuitively, one could guess that quantum 
theory could be of help in describing those decision-making processes that in some 
sense are of dual nature. For instance, when one needs to make a choice between the 
given alternatives, at the same time experiencing uncertainty with regard to these 
alternatives because of the existence of additional alternative attributes related 
to behavioral effects \cite{Yukalov_18,Yukalov_34,Yukalov_35,Yukalov_36}. 

A well known duality in decision making has to do with that being termed by different 
names, although expressing the same essence. This is the {\it cognition-emotion 
duality}, {\it rational-irrational duality}, {\it reasoning-affective duality, } or 
{\it conscious-subconscious duality}. There exists vast literature on the problem of 
studying the influence of emotions on the process of taking decisions, of which we 
mention just a few \cite{Leventhal_37,Schwarz_38,Pessoa_39,Scherer_40,Lerner_41,
Scherer_42,Reisenzein_43,Scherer_44}. In the mathematical language of quantum theory, 
this duality is described below \cite{Yukalov_18,Yukalov_34,Yukalov_35,
Yukalov_36,Yukalov_45,Yukalov_46}. 

In quantum measurement theory, the role of emotions is played by perturbations from 
surrounding, which can be called quantum noise. There are arguments \cite{Albrecht_66}
that, actually, all decisions and measurements are accompanied by a kind of quantum noise
due to random quantum fluctuations in the net of human neurons. The signal triggering the 
process of decision making travels along human neurons with an intrinsic temporal 
uncertainty of $1$ ms \cite{Faisal_67} accompanied by fluctuations in the number of open 
neuron ion channels. These molecular fluctuations are due to random Brownian motion of 
polypeptides in their surrounding fluid. Since the Brownian fluctuations are fundamentally 
quantum, the probabilities of events, in the strict sense, are also quantum.

The decision space  
$$
\cH = \cH_S \bigotimes \cH_A \; ,
$$
as in Sec. 6, is the tensor product of the subject space of mind and the space of 
alternatives. When choosing between the alternatives, the choice of the decision maker
includes two sides. From one side, one has to evaluate the utility of each alternative. 
But from the other side, the choice is accompanied by emotions associated with the 
given alternative. This is equivalent to saying that the decision maker considers not 
just the set of alternatives, but rather a set of composite prospects
\be
\label{82}
\pi_n \equiv A_n z_n \;  ,
\ee
each being formed by an alternative $A_n$ and the associated set of emotions represented 
by the notation $z_n$.

In quantum measurement theory, each process of measurement of an event $A_n$ is 
complimented by the accompanying quantum noise denoted through $z_n$. Again, from the 
mathematical point of view, the description is the same, whether we consider a measurement
accompanied by quantum noise or take a decision decorated by emotions.    

Emotions (noise) are set into correspondence to the vector
\be
\label{83}
|\; z_n \; \rgl = \sum_\al b_{n\al}\; | \; \al \; \rgl
\ee
pertaining to the subject space of mind $\cH_S$. The coefficients $b_{n\al}$ are random 
quantities, which reflects the randomness of emotions. These vectors do not need to be 
orthonormalized, so that the scalar product 
$$
\lgl\; z_m \; | \; z_n \; \rgl = \sum_\al b_{m\al}^* b_{n\al}
$$
is not necessarily the Kronecker delta. Respectively, the prospect states 
\be
\label{84}
|\; \pi_n \; \rgl = |\; A_n z_n \; \rgl = 
|\; A_n \; \rgl \; \bigotimes \; |\; z_n \; \rgl = 
\sum_\al b_{n\al} \; |\; A_n \al \; \rgl
\ee
also are not necessarily orthonormalized, 
$$
\lgl\; \pi_m \; | \; \pi_n \; \rgl =  \dlt_{mn} \; \lgl\; z_n \; | \; z_n \; \rgl \; .
$$

The prospect operator
\be
\label{85}
\hat P(\pi_n) = | \; \pi_n \; \rgl \lgl\; \pi_m \; | = 
| \; A_n z_n \; \rgl \lgl\; z_n  A_n \; |
\ee
is not idempotent, since
$$
\hat P(\pi_m) \; \hat P(\pi_n) = 
\dlt_{mn} \; \lgl\; z_n \; | \; z_n \; \rgl \; \hat P(\pi_n) \; .
$$
The matrix element
$$
\lgl\; \al A_m \; | \; \sum_k \hat P(\pi_k) \; | \; A_n \bt \; \rgl =
\dlt_{mn} b_{n\al}^* b_{n\bt}
$$
shows that the resolution of unity for the prospect operators cannot be satisfied in 
the strong sense, although can be valid in the weak sense of the average
\be
\label{86}
 \lgl \; \sum_n \hat P(\pi_n) \; \rgl = 
{\rm Tr}\; \hat\rho(t) \; \sum_n \; \hat P(\pi_n) = 1 \;  ,
\ee
where the trace is over the decision space $\mathcal{H}$. Thus the family
\be
\label{87}
\cP_\pi \equiv \{ \hat P(\pi_n) \}
\ee
of the prospect operators constitutes a positive operator-valued measure \cite{Yukalov_68}.
 
Therefore the triple 
\be
\label{88}
\{ \cH , \; \hat\rho(t) , \; \cP_\pi\}
\ee
forms the quantum probability space. The prospect probabilities 
\be
\label{89}  
 p(\pi,t) \equiv {\rm Tr}\; \hat\rho(t) \; \hat P(\pi_n) \;  ,
\ee
by the above construction, are semi-positive and normalized,
\be
\label{90}
 \sum_n \; p(\pi_n,t) = 1 \; , \qquad 0 \leq p(\pi_n,t) \leq 1 \; .
\ee

Substituting into the prospect probability (\ref{89}) the prospect operators (\ref{85})
makes it possible to separate in probability (\ref{89}) the non-negative part
\be
\label{91}
 f(\pi_n,t) \equiv \sum_\al \; | \; b_{n\al} \; |^2 \;
\lgl \; \al A_n \; | \; \hat\rho(t) \; | \; A_n \al \; \rgl \;  ,
\ee
which corresponds to classical probability, being normalized,
\be
\label{92}
 \sum_n \; f(\pi_n,t) = 1 \; , \qquad 0 \leq f(\pi_n,t) \leq 1 \;   .
\ee
The remaining part 
\be
\label{93}
q(\pi_n,t) \equiv \sum_{\al\neq\bt} \; b^*_{n\al} b_{n\bt} \; 
\lgl \; \al A_n \; | \; \hat\rho(t) \; | \; A_n \bt \; \rgl \; 
\ee
is not sign-defined, having the property
\be
\label{94}
 \sum_n \; q(\pi_n,t) = 1 \; , \qquad -1 \leq q(\pi_n,t) \leq 1 \;  .
\ee
In this way, the prospect probability (\ref{89}) is the sum of two terms:
\be
\label{95}
p(\pi_n,t) = f(\pi_n,t) + q(\pi_n,t) \;  .
\ee

The semi-positive term (\ref{91}) describes the classical probability or the 
fraction of decision makers choosing, between the given alternatives, following 
rational rules. It can be connected with the utility of alternatives by means of 
the Luce rule \cite{Luce_47} as follows. If the alternatives $A_n$ are characterized 
by the attributes $a_n$, then the probabilistic weight of each alternative is
\be
\label{96}
 f(\pi_n,t) = \frac{a_n}{\sum_n a_n} \qquad ( a_n \geq 0)  \; .
\ee
This term can be called {\it utility fraction} or {\it rational fraction}. For 
non-negative expected utilities, the natural attribute $a_n$ is the expected utility 
itself. Quantitative evaluation of this term has been studied in detail in Refs. 
\cite{Yukalov_18,Yukalov_46,Yukalov_48,Yukalov_49,Yukalov_50,Yukalov_51,Yukalov_52,
Favre_53,Yukalov_54}. 

The sign-undefined term (\ref{93}) is associated with emotional feelings of a decision 
maker inducing the attraction or repulsion towards the alternatives caused by emotional
evaluation of the quality of each alternative. Therefore this term can be named {\it 
attraction factor} or {\it quality factor}. By its definition, it is a random quantity,
being different for different subjects and even for the same subject at different times.
The type of randomness can be prescribed by a distribution over $q$. As is clear, 
being random, does not preclude to find its typical average value that plays 
the role of a non-informative prior. The non-informative prior for the attraction factor 
was found \cite{Yukalov_18,Yukalov_45,Yukalov_46,Yukalov_49,Yukalov_51,Yukalov_52,
Yukalov_54} to be $\pm 0.25$, with the sign defined by attractiveness or repulsiveness 
of the corresponding prospect. 

In the process of quantum measurements, the term $q(\pi_n,t)$ is caused by random noise. 
In that case, one needs to define a distribution over $q$. If the properties of the noise
are characterized by average values of $q$, these values will define the typical uncertainty
of a measurement.    

Summarizing, the prospect probability, taking account of both the utility of 
alternatives, as well as their attractiveness, can be evaluated by the formula
\be
\label{97}
p(\pi_n,t) = f(\pi_n,t) \pm 0.25 \;  ,
\ee
provided the normalization condition (\ref{90}) holds true.   
 
In the references cited above, there are numerous examples of the use of the described
approach to decision making, demonstrating good numerical agreement with empirical data.
Here we give just one example showing how this approach allows for the explanation of 
the decoy effect \cite{Yukalov_52}.

\section{Example of Decoy Effect}

The decoy effect was described by Huber et al. \cite{Huber_55} and observed in many
experimental studies, e.g. 
\cite{Simonson_56,Wedell_57,Tversky_58,Ariely_59,Mao_60,Chernev_61}. The decoy effect 
describes how, when we are choosing between two alternatives, the addition of a third, 
less attractive option (the decoy) can influence our perception of the original two 
choices. Decoys are completely inferior to one option (the target) but only partially 
inferior to the other (the competitor). For this reason, the decoy effect is sometimes 
called the asymmetric dominance effect. 

The meaning of the decoy effect can be illustrated by the following example. Suppose a 
buyer is choosing between two objects, $A$ and $B$. The object $A$ is of better quality, 
but of higher price, while the object $B$ is of slightly lower quality, while less 
expensive. As far as the functional properties of both objects are close to each other, 
but $B$ is cheaper, the majority of buyers prefer the object $B$. But then the sales 
person slightly changes the situation presenting a third object $C$, which is of about 
the same quality as $A$, but of a much higher price than $A$. This causes the buyer to 
reconsider the choice between the objects $A$ and $B$, while the object $C$, having 
practically the same quality as $A$ but being much more expensive, is of no interest. 
Choosing now between $A$ and $B$, the majority of buyers prefer the higher quality but 
more expensive object $A$. The object $C$, being not a choice alternative, plays the 
role of a decoy. Experimental studies confirmed the decoy effect for a variety of 
objects: cars, microwave ovens, shoes, computers, bicycles, beer, apartments, and many 
others.

The decoy effect contradicts the regularity axiom in decision making telling that if 
$B$ is preferred to $A$ in the absence of $C$, then this preference has to remain in 
the presence of $C$. From the point of view of classical decision theory, the decoy 
effect is a paradox. Decoys work subconsciously, being an example of a behavioral nudge 
that steers individuals towards making a certain choice. As other subconscious and 
emotional influences in decision making, this effect can be explained in the frame of 
the approach described in the previous Sec. 10. 

Assume that buyers are choosing between two objects A and B, say between two ovens, 
as in the experiment of Ariely and Wallsten \cite{Ariely_59}. The oven $A$ is of higher 
quality but more expensive, while the oven $B$ is of moderate quality but cheaper. 
Since the quality of these two ovens is not much different, but $B$ is cheaper, the 
majority prefer the oven $B$. In the experiment \cite{Ariely_59}, the fraction of 
buyers preferring either $A$ or $B$ was
$$
 f(A) = 0.4 \; , \qquad  f(B) = 0.6 \;  .
$$
Then the third oven $C$ was displayed costing much more expensive than $A$, but being 
of just slightly higher quality. The oven $C$, being of much higher price than $A$, 
but of close quality, is not a choice. However, it attracted attention of buyers to 
the quality of the ovens. The emotional attraction of buyers induces the attraction 
factors $q(A)$ and $q(B)$. In that sense, we need to consider not simply alternatives 
$A$ and $B$, but the prospects $\pi_A=A z_A$ and $\pi_B=B z_B$, although, for the 
simplicity of notation, we continue writing just $A$ and $B$. As far as the feature 
of quality has become more attractive, the related attraction factors are such that 
$q(A)>q(B)$. Estimating them as non-informative priors, we have $q(A)=0.25$ and 
$q(B)=-0.25$. Therefore the theory predicts the choice probabilities 
$$
 p(A) = f(A) + q(A) = 0.4 + 0.25 = 0.65 \; ,  \qquad
 p(B) = f(B) + q(B) = 0.6 - 0.25 = 0.35 \; .
$$
In other words, we obtain the estimates predicting the real fraction of buyers preferring 
this or that oven. Comparing this prediction with the experimental data
$$
  p_{exp}(A) = 0.61 \;  , \qquad  p_{exp}(B) = 0.31 \;  ,
$$
we see that they are very close to each other.

\section{Conclusions}

The central point of the paper is the problem of defining conditional probability 
in quantum theory. In order to clearly elucidate the difficulties and possible ways 
of constructing quantum conditional probabilities, the consideration has been based 
on the example of tossing quantum coins and dice. This case is of interest by its 
own, since quantum coins and dice present a paradigm of quantum bits that are widely 
employed in quantum information processing, quantum computing, and quantum decision 
making.  

There exist two types of quantum conditional probabilities that can be called a
temporal conditional probability and a spatial conditional probability. The temporal 
conditional probability considers two events occurring at different times. It is based 
on the quantum state reduction after a measurement and uses the von-Neumann-L\"{u}ders 
state. The latter is the probability $p(B_k,t_0+0|A_n,t_0)$ of an event, say $B_k$, at 
the time $t_0+0$, occurring immediately after another event, say $A_n$, has certainly 
occurred at the previous time $t_0<t_0 + 0$. It is shown that, strictly speaking, the 
L\"{u}ders probability cannot be reduced to the classical conditional probability. 
Therefore it cannot be treated as a generalization of the latter. The L\"{u}ders 
probability and the classical conditional probability are different notions valid 
for different systems, either quantum or classical. The key difference between these 
probabilities is caused by the state reduction taking place for quantum measurements, 
but being absent for classical systems. The direct consequence of the state reduction 
is the principle of measurement reproducibility in quantum theory \cite{Neumann_27} 
according to which the second measurement, accomplished immediately after the first 
one, has to reproduce the result of the latter. No such principle exists for classical 
systems.

When one attempts to apply quantum techniques to cognitive sciences, with using the 
L\"{u}ders probability, one should take into account a decision maker and the related 
subject state of mind. Then, although the system state experiences the reduction in 
the space of alternatives, but it remains more or less invariant in the subject space, 
thus treating the decision maker as a self-consistent personality.    
  
Another type of probability is the spatial conditional quantum probability 
$p(A_n|B_k,t)$ that considers two events occurring simultaneously in time, but in 
different spatial locations. The probability $p(A_n|B_k,t)$ is the probability of 
an event $A_n$ occurring in one location at time $t$, under the condition that at 
the same time the other event $B_k$ occurs in another location. The structure and 
properties of this probability are close to classical conditional probability.    

The dual nature of quantum theory allows to extend the classical decision making 
to the problems where behavioral effects are of great importance. In that case it 
is necessary to take account of the subject space of mind and to consider not 
merely alternatives but composite prospects of alternatives decorated by emotions 
accompanying the latter. The suggested approach of taking into account emotions 
yields good agreement of predicted and experimentally observed data.  

In the process of quantum measurements, the role of emotions is played by quantum 
fluctuations accompanying the measurements. At the same time, the origin of emotions,
to some extent, can be due to quantum fluctuations in the brain. The existence of 
such fluctuations has been justified by physiological studies demonstrating the 
presence of Brownian fluctuations in the neuron net.    

\section*{Acknowledgments}

The author is grateful for useful discussions to J. Harding, H. Nguyen, D. Sornette, 
and E.P. Yukalova.

\end{document}